\documentclass[10pt,twocolumn,letterpaper]{article}

\usepackage{wacv}              

\usepackage{graphicx}
\usepackage{amsmath}
\usepackage{amssymb}
\usepackage{booktabs}
\usepackage{multirow}
\usepackage[table,xcdraw]{xcolor}
\usepackage[normalem]{ulem}
\useunder{\uline}{\ul}{}
\usepackage{adjustbox}

\usepackage[pagebackref,breaklinks,colorlinks]{hyperref}

\usepackage[capitalize]{cleveref}
\crefname{section}{Sec.}{Secs.}
\Crefname{section}{Section}{Sections}
\Crefname{table}{Table}{Tables}
\crefname{table}{Tab.}{Tabs.}

\def\wacvPaperID{265} 
\def\confName{WACV}
\def\confYear{2025}

\begin{document}

\title{Boosting Diffusion Guidance via Learning Degradation-Aware Models \\ for Blind Super Resolution} 



\author{Shao-Hao Lu$^{1}$ \qquad Ren Wang$^{2}$ \qquad Ching-Chun Huang $^{1}$ \qquad Wei-Chen Chiu$^{1}$ 
\\
{\small\tt $^{1}$National Yang Ming Chiao Tung University \qquad $^{2}$MediaTek Inc.}
\\
{\small\tt \{shlu2240.cs11, chingchun, walon\}@nycu.edu.tw \qquad ren.wang@mediatek.com}}

\maketitle

\definecolor{gold}{RGB}{241, 154, 33}
\newcommand{\ren}[1]{#1}
\newcommand{\rencam}[1]{#1}
\newcommand{\shlu}[1]{#1}
\newcommand{\walon}[1]{#1}

\newcommand{\heading}[1]{\vspace{1mm}\noindent\textbf{#1}}

\begin{abstract}
\rencam{Recently, diffusion-based blind super-resolution (SR) methods have shown great ability to generate high-resolution images with abundant high-frequency detail, but the detail is often achieved at the expense of fidelity. Meanwhile, another line of research focusing on rectifying the reverse process of diffusion models (i.e., diffusion guidance), has demonstrated the power to generate high-fidelity results for non-blind SR. However, these methods rely on known degradation kernels, making them difficult to apply to blind SR. To address these issues, we present DADiff in this paper. DADiff incorporates degradation-aware models into the diffusion guidance framework, eliminating the need to know degradation kernels. Additionally, we propose two novel techniques---input perturbation and guidance scalar---to further improve our performance. Extensive experimental results show that our proposed method has superior performance over state-of-the-art methods on blind SR benchmarks.}

\end{abstract}

\section{Introduction}
Single image blind super-resolution (blind-SR) aims to recover a high-resolution (HR) image from a low-resolution (LR) observation which undergoes an unknown degradation process, where the loss of high-frequency detail during the unknown degradation primarily leads to the inherent challenge in the task of blind-SR. \rencam{Another} well-known challenge comes from having multiple potential solutions of high-resolution images (i.e., multiple HR images would yield the identical LR images)\rencam{, and} thus making the ill-posed nature of super-resolution problem.
Although the renaissance of deep learning techniques in recent decade brings the magic leap to the task of super-resolution, especially the common practice stemmed from regression-based methods \cite{lim2017enhanced, li2022learning} excels in fidelity, they often suffer from \rencam{insufficient high-frequency image detail}.  
This deficiency highlights the need for generative models, where the essence lies in drawing high-frequency detail from the image prior distribution captured by generative models. Previous approaches have turned to Generative Adversarial Networks (GANs) \cite{goodfellow2014generative, zhang2021designing, wang2021real} for such purpose, yet there comes other issues \rencam{such as unstable training or mode collapse}.

With the recent advances in diffusion-based generative models, some works \cite{rombach2022high, wang2023exploiting, yue2024resshift, yang2023pixel, wu2024seesr, lin2023diffbir} have utilized pretrained diffusion models to address the blind super-resolution problem for better training stability, image synthesis quality, and more flexible controls of generative procedure.
\rencam{Specifically, some works~\cite{wang2023exploiting, yue2024resshift, lin2023diffbir}} propose to \rencam{directly} guide the diffusion sampling with LR input, while \rencam{some works~\cite{yang2023pixel, wu2024seesr} use additional prompts for guidance}. Though these methods can produce high-quality images, they often fall short in terms of fidelity.
Several non-blind diffusion-based image restoration methods \cite{kawar2022denoising, wang2022zero} have attempted to address this challenge, demonstrating the ability to produce \rencam{high-fidelity results}.
However, these methods have limitations in that the degradation operator must be \rencam{both known and linear, whereas degradation operators are typically unknown in real-world scenarios}.

In this work, we \rencam{adopt the Denoising Diffusion Null-Space Model (DDNM)~\cite{wang2022zero}, a representative non-blind diffusion-based image restoration method, as our backbone framework. In order to eliminate the need to know degradation kernels in DDNM, we introduce several modifications.}
\rencam{First, we use a deep neural network to approximate the degradation process, eliminating the need to provide the degradation operator. This network is called degradation model.} The degradation model not only addresses unknown degradation kernels, but also potentially overcomes the limitation of kernel linearity, enhancing the versatility and effectiveness of our proposed method. \rencam{Second, for the inverse matrix of the degradation operator in DDNM \walon{(noting that since degradation operator in DDNM is represented as matrix, its inverse is thus related to the degradation removal)}, we replace it with a restoration model. During the sampling process, the restoration model aims to recover an HR image from an LR image generated by our degradation model at each timestep. Last, to improve performance, both of the degradation model and the restoration model are conditioned on the \walon{same} degradation representations, which are extracted from the LR input by an encoder model. As a result, we call these two models degradation-aware models.}

 Moreover, \rencam{we propose two additional} techniques to enhance the capacity and performance of our method against the prediction errors upon the degradation: 1) \rencam{\emph{input perturbation} is introduced} to reduce the performance drops associated with the inaccuracies in degradation estimation, and thus contributing to model stability; 2) \rencam{\emph{guidance scalar}} is introduced to facilitate better integration between the restoration model and the pretrained diffusion model as combining the restoration model with diffusion sampling guidance could alter the data distribution predefined by the diffusion model\rencam{, which leads} to issues such as hallucinations and artifacts.
 With all the \rencam{above-mentioned} components, \rencam{we present a robust method} for the task of diffusion-based blind super-resolution, \rencam{and} extensive experiments demonstrate that our method has the capability to generate realistic details without sacrificing the fidelity. Compared to other state-of-the-art generative \rencam{model}-based blind-SR methods, our approach excels in both fidelity and \rencam{perceptual quality}.
 
 Our contributions are \rencam{summarized} as follows:

\begin{itemize}
    \item \rencam{We propose degradation-aware models, which consists of a degradation model and a restoration model. With these two models, we eliminate the need to know degradation kernels in DDNM~\cite{wang2022zero}, making it applicable to blind-SR.}
    \item \rencam{We propose input perturbation and guidance scalar, \walon{ which contribute to make our method more tolerant towards} the inaccuracies in degradation estimation.}
    \item \rencam{Extensive experiments demonstrate that our method has superior performance over other state-of-the-art generative model-based blind-SR methods, in terms of both fidelity and perceptual quality.}
\end{itemize}

\section{Related Works}
\subsection{Image Super-Resolution}

Image \rencam{super-resolution} (SR) aims to restore a high-resolution (HR) image from its degraded low-resolution (LR) observation. Early methods \cite{chen2021pre, dong2014learning, dong2015image, lim2017enhanced} \rencam{usually} assume a predefined degradation \rencam{operator such as bicubic downsampling. These methods often fail in real-world data \walon{while} the degradation assumption is not accurate.}
\rencam{Therefore,} recent works have shifted their focus \rencam{towards} blind-SR, where the degradation is unknown. \shlu{Some methods \cite{wang2021unsupervised, maeda2020unpaired, zhang2021blind, zhu2017unpaired, li2022learning} use unsupervised learning techniques to learn degradation models, while others \cite{wang2021real, zhang2021designing, zhou2022towards, luo2024and} aim to synthesize LR-HR image pairs that resemble real-world data. However, capturing diverse real-world degradations remains challenging.}

\subsection{Diffusion-based Image Super-Resolution}
Recent advances in diffusion models \cite{ho2020denoising, nichol2021glide, song2020score, rombach2022high} have shown promise in image synthesis tasks. \rencam{Many image restoration (IR) methods \cite{chung2023parallel, fei2023generative, kawar2022denoising, wang2022zero} use diffusion models as an image prior. Some non-blind IR methods \cite{kawar2022denoising, wang2022zero} demonstrate the ability to produce high-fidelity results with known degradation operators. Some works~\cite{chung2023parallel, fei2023generative} extend non-blind IR to blind IR by estimating the degradation operator, and thus knowing the degradation operator becomes unnecessary.} However, \rencam{all of these methods} require the degradation operator to be \rencam{linear, which should be over-simplified in real-world scenarios.}

\rencam{Most diffusion-based blind super-resolution methods \cite{choi2021ilvr, yue2024resshift, wang2023exploiting, wu2024seesr, lin2023diffbir, yang2023pixel} do not need a known or estimated degradation operator. Instead, they usually use conditioning signals to guide diffusion models during sampling.} Stable-SR \cite{wang2023exploiting} and Diff-BIR \cite{lin2023diffbir} utilize only the low-resolution observation as \rencam{the conditioning signal.} DASR \cite{yang2023pixel} and SeeSR \cite{wu2024seesr} further incorporate semantic information as \rencam{an extra} conditioning signal. All of these methods have shown highly promising capabilities in generating realistic image detail. However, they often fall short in terms of fidelity due to the inherent diffusion hallucination.  Achieving high-fidelity results remains a significant challenge for \rencam{most diffusion-based blind-SR methods}.

\section{Preliminaries}
\subsection{Diffusion Models}
We follow the diffusion model defined in DDPM \cite{ho2020denoising}.
\ren{Given a natural image $\mathbf{x}_0 \sim q(\mathbf{x})$, DDPM uses} a forward Markovian diffusion process
\ren{$ q(\mathbf{x}_t | \mathbf{x}_{t-1}) $ to repeatedly add Gaussian noise on it until $t=T$ by}
\begin{align}
q(\mathbf{x}_{t}|\mathbf{x}_{t-1}) = \mathcal N(\mathbf{x}_{t};\sqrt{1-\beta_{t}}\mathbf{x}_{t-1},\beta_{t}\mathbf{I}),
\end{align}
\ren{where $\beta_t$ is the pre-defined variance at timestep $t$. Note that if $T$ is sufficiently large (e.g., $1000$), we can have $\mathbf{x}_T \sim \mathcal{N}(\mathbf{0}, \mathbf{I})$.}
Based on the property of Markov chain, for any intermediate timestep \(t \in \{1, 2, ..., T\}\), the corresponding noisy distribution has an analytic form:
\begin{align}
q(\mathbf{x}_{t}|\mathbf{x}_{0}) &= \mathcal N(\mathbf{x}_{t};\sqrt{\overline{\alpha}_{t}}\mathbf{x}_{0},(1-\overline{\alpha}_{t})\mathbf{I}) \nonumber \\
        &= \sqrt{\overline{\alpha}_{t}}\mathbf{x}_{0} + \sqrt{1-\overline{\alpha}_{t}}\mathbf{\epsilon},
\end{align}
where \(\overline{\alpha}_{t} = \prod_{i}^{t}(1-\beta_{i})\) and \(\mathbf{\epsilon} \sim \mathcal N(\mathbf{0}, \mathbf{I})\).
The reverse process aims at yielding the previous state $\mathbf{x}_{t-1}$ from $\mathbf{x}_{t}$ using the posterior distribution:
\begin{align}
p(\mathbf{x}_{t-1}|\mathbf{x}_{t}, \mathbf{x}_{0|t})=\mathcal N(\mathbf{x}_{t-1};\mathbf{\mu}_{t}(\mathbf{x}_{t}, \mathbf{x}_{0|t}), \sigma_{t}^{2}\mathbf{I}),
\label{eq:posterior}
\end{align}
\ren{where
\begin{align}
\mathbf{\mu}_t(\mathbf{x}_t, \mathbf{x}_{0|t}) &= \frac{\sqrt{\overline{\alpha}_{t-1}}\beta_t}{1-\overline{\alpha}_t} \mathbf{x}_{0|t} + \frac{\sqrt{\alpha_t}(1-\overline{\alpha}_{t-1})}{1-\overline{\alpha}_t}\mathbf{x}_t,
\label{eq:mu} \\
\sigma_t^2 &= \frac{1-\overline{\alpha}_{t-1}}{1-\overline{\alpha}_t} \beta_t,
\end{align}
and
\begin{align}
\mathbf{x}_{0|t} = \frac{1}{\sqrt{\overline{\alpha}_t}} \left( \mathbf{x}_t - \sqrt{1-\overline{\alpha}_t} \mathbf{\epsilon}_\theta(\mathbf{x}_t, t) \right),
\label{eq:x0}
\end{align}
where $\mathbf{\epsilon}_\theta(\mathbf{x}_t, t)$ is the estimated noise $\mathbf{\epsilon}$ at timestep $t$. By iteratively sampling \(\mathbf{x}_{t-1}\) from \(p(\mathbf{x}_{t-1}|\mathbf{x}_{t}, \mathbf{x}_{0|t})\), DDPM can generate realistic image \(\mathbf{x}_{0} \sim q(\mathbf{x})\) from random noise \(\mathbf{x}_{T} \sim \mathcal N(\mathbf{0}, \mathbf{I})\).}


\subsection{Denoising Diffusion Null-Space Model}
\rencam{Denoising Diffusion Null-Space Model (DDNM)} \cite{wang2022zero}
offers a zero-shot framework for solving a range of arbitrary linear inverse problems. This innovative approach leverages a pre-trained off-the-shelf diffusion model as a generative prior, requiring no additional training or modifications to existing networks. By focusing solely on refining the null-space contents during the reverse diffusion process, DDNM has the capability to generate diverse results that satisfy both \rencam{fidelity and perceptual quality}.

\ren{Given a degraded image $\mathbf{y} \in \mathbb{R}^{d \times 1}$ and a degradation operator $\mathbf{A} \in \mathbb{R}^{d \times D}$, we could have $\mathbf{y}=\mathbf{A}\mathbf{x}$, where $\mathbf{x} \in \mathbb{R}^{D \times 1}$ is the ground-truth image. DDNM applies this relationship to rectify the $\mathbf{x}_{0|t}$ in \cref{eq:x0} as}
\begin{align}
\hat{\mathbf{x}}_{0|t} = \mathbf{A}^{\dagger}\mathbf{y} + (\mathbf{I} - \mathbf{A}^{\dagger}\mathbf{A}) \mathbf{x}_{0|t},
\label{eq:f2}
\end{align}
where \(\hat{\mathbf{x}}_{0|t}\) can be seen as the composition of $\mathbf{x}_{0|t}$ and $\mathbf{y}$ through range-null space decomposition. \ren{Then, they substitute the $\mathbf{x}_{0|t}$ in \cref{eq:mu} by this rectified version and sample $\mathbf{x}_{t-1}$ using \cref{eq:posterior}.}


The strength of DDNM lies in its ability to utilize the denoising diffusion model to fill in the null-space information, ensuring that the output adheres to the notion of \textit{Realness}. This adherence is particularly effective in scenarios where the degradation is linear and known, leading to perfect \textit{Consistency} due to the natural properties of the Range-Null space. However, a notable limitation arises when the degradation operator \(\mathbf{A}\) must be both known and linear, which
\ren{is often not aligned} with the complexities of real-world applications where degradations are typically unknown.

\begin{figure*}
    \centering
    \includegraphics[width=\linewidth]{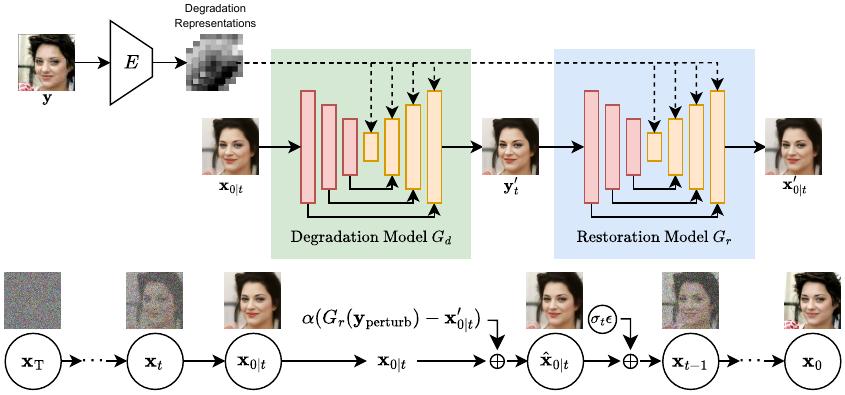}
    \caption{\heading{Overview of our method.} At each timestep $t$ of a reverse process, given the output of a pretrained diffusion model $\mathbf{x}_{0|t}$ and a degraded image $\mathbf{y}$, our degradation model $G_d$ takes as input $\mathbf{x}_{0|t}$ and degradation representations $E(\mathbf{y})$, and produces an estimated degraded image $\mathbf{y}'_t$. Then, our restoration model $G_r$ transforms $\mathbf{y}'_t$ into a restored image $\mathbf{x}'_{0|t}$. We then rectify $\mathbf{x}_{0|t}$ to $\hat{\mathbf{x}}_{0|t}$ using $\mathbf{x}'_{0|t}$ and $G_r(\mathbf{y}_\text{perturb})$, where $\mathbf{y}_\text{perturb}$ is a perturbed degraded image, and add noise $\sigma_t \epsilon$ to obtain $\mathbf{x}_{t-1}$ for the next timestep. Repeating this process, we get the final restored image $\mathbf{x}_0$ at timestep $0$.}
    \label{fig:implicit}
\end{figure*}

\section{Method}
We follow the Denoising Diffusion Null-Space Model (DDNM)~\cite{wang2022zero} framework, considering its capability to generate high-quality images while preserving high fidelity. To run \ren{the} DDNM algorithm, the first step involves obtaining the degradation operator \ren{$\mathbf{A}$} and its corresponding pseudo inverse \ren{$\mathbf{A}^\dagger$}. We investigate \rencam{replacing the degradation operator $\mathbf{A}$ and its corresponding pseudo-inverse $\mathbf{A}^\dagger$ with a DNN-based degradation model and a DNN-based restoration model, both of which are in a more general form}. 
The purpose of the degradation model is to imitate the degradation process. \rencam{During sampling, the restoration model will recover an HR image from an LR image generated by the degradation model at each timestep. Furthermore, to improve performance, both of the degradation model and the restoration model will \rencam{be conditioned on \walon{the same} degradation representations} extracted from the LR input \rencam{by an encoder model}. In \cref{sec:implicit}, we give a detailed formulation for all of these models. In \cref{sec:perturbY}, we present \emph{input perturbation}, which aims to mitigate performance drops caused by inaccuracies in degradation estimation. In \cref{sec:scalar}, we present \emph{guidance scalar}, which aims to facilitate better integration between the
restoration model and DDNM to avoid hallucinations and artifacts}.

\subsection{Degradation-Aware Models}
\label{sec:implicit}
Li \etal \cite{li2022learning} introduce a method to encode latent degradation representations from degraded images through the joint learning of image degradation and restoration. The process begins with employing an encoder \(E\) to encode the degraded image \ren{$\mathbf{y}$} into the degradation representations \ren{$E(\mathbf{y})$}. This representation is modeled as a multi-channel latent map, capturing the 2D spatially varying degradation in a latent space. Subsequently, the framework introduces an image degradation model \(G_{d}\) and an image restoration model \(G_{r}\) to produce the degraded image \ren{$\mathbf{y}'$} and the restored image \ren{$\mathbf{x}'$}, respectively. The entire framework is trained in an end-to-end manner.

In the context of our work,
\ren{we replace the $\mathbf{A}$ and $\mathbf{A}^\dagger$ in DDNM with $G_d$ and $G_r$,} \rencam{with the aim of removing the limitation on known and linear degradation kernels.
If we substitute the $\mathbf{A}$ and $\mathbf{A}^\dagger$ in \cref{eq:f2} by $G_d$ and $G_r$, we can rewrite \cref{eq:f2} as}
\begin{align}
\begin{split}
\hat{\mathbf{x}}_{0|t} = \mathbf{x}_{0|t} + (G_{r}(\mathbf{y}) - \mathbf{x}_{0|t}'),
\label{eq:main_eq}
\end{split}
\end{align}
\rencam{where $\mathbf{x}_{0|t}' = G_{r}(G_{d}(\mathbf{x}_{0|t}))$. Note that \cref{eq:main_eq} can be interpreted as guiding $\mathbf{x}_{0|t}$ towards $G_{r}(\mathbf{y})$.}

Notably, the supervision for \(G_{d}\) and \(G_{r}\) in \ren{this} framework are separated, indicating the independence of these two models. To ensure the pseudo-inverse relationship between $G_d$ and $G_r$, we introduce a consistency loss during the joint training of \(E, G_{d}, G_{r}\) as 
\begin{equation}
\mathcal{L}_c = \| \mathbf{y} - G_{d}(G_{r}(\mathbf{y}, E(\mathbf{y})), E(\mathbf{y})) \|_{1}.
\label{eq:consistencyLoss}
\end{equation}
This additional loss term promotes stability and accelerates convergence during training.

\rencam{It is expected that replacing $\mathbf{A}$ and $\mathbf{A}^\dagger$ with neural networks could easily yield promising results.}
However, due to the inherent estimation errors and the potential misalignment between \ren{$G_d$} and \ren{$G_r$}, the Range-Null space decomposition in DDNM may not perfectly hold.
To address \rencam{this issue, we present input perturbation and guidance scalar in the next sections.} These refinements aim to further enhance the synergy between DDNM and the \rencam{degradation-aware models $G_d$ and $G_r$, and thereby address potential issues caused by degradation inconsistencies to improve} the overall performance of our method.

\begin{table*}[ht!]
\centering
\begin{adjustbox}{max width=\textwidth}
\begin{tabular}{c|ccccccccccccc}
\hline
\multicolumn{1}{l|}{Metric} & MsdiNet \cite{li2022learning}                        & BSRGAN \cite{zhang2021designing}  & \begin{tabular}[c]{@{}c@{}}Real-\\ ESRGAN\end{tabular} \cite{wang2021real} & LDL \cite{liang2022details}     & DASR \cite{liang2022efficient}    & FeMaSR \cite{chen2022real}  & DDNM \cite{wang2022zero}     & LDM \cite{rombach2022high}                            & ResShift \cite{yue2024resshift}                      & PASD \cite{yang2023pixel}    & DiffBIR \cite{lin2023diffbir}  & SeeSR \cite{wu2024seesr}   & Ours                           \\ \hline
PSNR ↑                       & {\color[HTML]{FF0000} 27.3933} & 22.0694 & 21.7337                                                & 23.1245 & 22.9961 & 20.2219 & 23.0214  & 21.7728                        & 21.6208                       & 22.7452 & 20.7647  & 21.8206 & {\color[HTML]{0000FF} 24.4681} \\
SSIM ↑                       & {\color[HTML]{FF0000} 0.7276}  & 0.5188  & 0.5179                                                 & 0.5355  & 0.5369  & 0.4601  & 0.5459   & 0.5108                         & 0.5101                        & 0.5348  & 0.4422   & 0.5009  & {\color[HTML]{0000FF} 0.6208}  \\
LPIPS ↓                      & 0.3486                         & 0.3392  & 0.3486                                                 & 0.3498  & 0.3682  & 0.3242  & 0.4814   & 0.3238                         & {\color[HTML]{0000FF} 0.3097} & 0.412   & 0.4106   & 0.3399  & {\color[HTML]{FF0000} 0.3069}  \\
DISTS ↓                      & 0.2558                         & 0.2634  & 0.2681                                                 & 0.2746  & 0.2824  & 0.2573  & 0.3182   & {\color[HTML]{0000FF} 0.2474}  & 0.2543                        & 0.2855  & 0.2866   & 0.2599  & {\color[HTML]{FF0000} 0.2393}  \\
FID ↓                        & 91.772                         & 92.604  & 91.3774                                                & 94.2726 & 102.948 & 94.2519 & 106.1603 & {\color[HTML]{0000FF} 83.9125} & 88.3887                       & 90.0368 & 101.1623 & 86.939  & {\color[HTML]{FF0000} 82.2803} \\ \hline
\end{tabular}
\end{adjustbox}
\caption{\heading{Quantitative comparison of $\bf{4 \times}$ upsampling on \emph{DIV2K-Val}.} The best performance is in \textcolor{red}{red} while the second best is in \textcolor{blue}{blue}.}
\label{tab:div2k}
\end{table*}

\begin{table*}[]
\begin{adjustbox}{max width=\textwidth}
\begin{tabular}{c|ccccc|ccccc}
\hline
                         & \multicolumn{5}{c|}{$4 \times$}                                                                                                                                         & \multicolumn{5}{c}{$8 \times$}                                                                                                                                          \\ 
\multirow{-2}{*}{Method} & PSNR ↑                         & SSIM ↑                        & LPIPS ↓                       & DISTS ↓                       & FID ↓                          & PSNR ↑                         & SSIM ↑                        & LPIPS ↓                       & DISTS ↓                       & FID ↓                          \\ \hline
DDNM \cite{wang2022zero}    & 25.6975                        & {\color[HTML]{0000FF} 0.8069} & 0.2074                        & 0.1634                        & 38.3359                        & 20.1491                        & 0.65                          & 0.2469                        & 0.2061                        & 52.6844                        \\
SR3 \cite{saharia2022image}     & 26.0082                        & 0.7392                        & {\color[HTML]{0000FF} 0.1289} & {\color[HTML]{0000FF} 0.1209} & {\color[HTML]{0000FF} 19.4323} &  {\color[HTML]{0000FF} 23.79}                          & {\color[HTML]{0000FF} 0.6995} & {\color[HTML]{0000FF} 0.1742} & {\color[HTML]{0000FF} 0.1505} & {\color[HTML]{0000FF} 25.6039} \\
DiffBIR \cite{lin2023diffbir} & 22.9068                        & 0.7098                        & 0.1769                        & 0.1793                        & 66.8081                        & 19.0676                        & 0.5621                        & 0.264                         & 0.2102                        & 67.7493                        \\
SeeSR \cite{wu2024seesr}   & 22.3977                        & 0.7187                        & 0.1808                        & 0.1747                        & 66.7649                        & 19.2302                        & 0.5827                        & 0.2542                        & 0.1974                        & 67.6858                        \\
MsdiNet \cite{li2022learning} & {\color[HTML]{FF0000} 28.8877} & {\color[HTML]{FF0000} 0.8317} & 0.2405                        & 0.2048                        & 59.757                         & {\color[HTML]{FF0000} 25.2789} & {\color[HTML]{FF0000} 0.7527} & 0.3426                        & 0.2555                        & 76.6328                        \\
Ours    & {\color[HTML]{0000FF} 27.1513} & 0.7784                        & {\color[HTML]{FF0000} 0.0913} & {\color[HTML]{FF0000} 0.1066} & {\color[HTML]{FF0000} 18.6689} & 22.8065 & 0.6893                        & {\color[HTML]{FF0000} 0.1502} & {\color[HTML]{FF0000} 0.1399} & {\color[HTML]{FF0000} 23.9716} \\ \hline
\end{tabular}
\end{adjustbox}
\caption{\heading{Quantitative comparison on \emph{CelebA-Val}.} The best performance is in \textcolor{red}{red} while the second best is in \textcolor{blue}{blue}.}
\label{tab:quantitative}
\end{table*}

\subsection{Input Perturbation}
\label{sec:perturbY}

\shlu{In \cref{eq:main_eq}, $\mathbf{y}$ represents the ground-truth degraded image, \rencam{which can be though of as obtained by $\mathbf{y}=G_d^{\text{gt}}(\mathbf{x})$, where $G_d^{\text{gt}}$ is the ground-truth degradation operator and $\mathbf{x}$ is the ground-truth high-resolution image. As a result, $G_r(\mathbf{y})$ and $G_r(G_d(\mathbf{x}_{0|t}))$ cooperate with different degradation operators, and this mismatch could lead to failure cases.} Intuitively, we attempt to \rencam{perform a perturbation on $\mathbf{y}$ such that it were also obtained by applying $G_d$ as the degradation operator.}}


However, directly transforming \shlu{$\mathbf{y}$ into $G_d(\mathbf{x})$} 
is infeasible due to the unknown ground-truth \rencam{high-resolution} image \ren{$\mathbf{x}$}. To address this, we propose a perturbation approach using
\shlu{$G_d$ and $G_r$ sequentially as}
\begin{equation}
\shlu{\mathbf{y}_\text{perturb} = G_{d}(G_{r}(\mathbf{y})).}
\label{eq:perturby}
\end{equation}
\rencam{If we substitute the $\mathbf{y}$ in \cref{eq:main_eq} by $\mathbf{y}_\text{perturb}$, the degradation operators in \cref{eq:main_eq} become aligned}.
The choice of \shlu{$G_r$} is flexible and can be any off-the-shelf restoration model.
\shlu{In our method, we can simply use the $G_r$ described in \cref{sec:implicit}.}
By applying $\mathbf{y}_\text{perturb}$ to \shlu{\cref{eq:main_eq}}, \rencam{our method is expected to become more robust}.


\subsection{Guidance Scalar}
\label{sec:scalar}

Note that $G_r(\mathbf{y})$ represents the output of a restoration model. \ren{We observe that applying $G_r(\mathbf{y})$ to \shlu{\cref{eq:main_eq}} significantly accelerates the restoration process. This might bring a negative effect on fidelity as the diffusion model would take more time to hallucinate nonexistent contents.} 
To counteract this, it becomes imperative to reduce the influence of
\ren{$G_r(\mathbf{y})$} in \shlu{\cref{eq:main_eq}}, slowing down the restoration process to find an optimal balance.
\ren{That is, we should be able to control the restoration strength during the diffusion process.}

\rencam{Recall that a characteristic of \shlu{\cref{eq:main_eq}} is that it} can be viewed as guiding \ren{$\mathbf{x}_{0|t}$} towards \shlu{$G_{r}(\mathbf{y})$}.
However, in this case, the approximation error from \rencam{$G_d(\mathbf{x}_{0|t})$} will make the update direction not correctly point towards \rencam{$G_r(\mathbf{y})$}. As a result, how to determine an appropriate weighting of this guidance is also important from this perspective.



To this end, we introduce a scalar $\alpha \in [0, 1]$ to rewrite \shlu{\cref{eq:main_eq} as}
\begin{align}
\hat{\mathbf{x}}_{0|t} = \mathbf{x}_{0|t} + \alpha (G_{r}(\mathbf{y}) - \mathbf{x}_{0|t}' ),
\label{eq:alpha_guidance}
\end{align}
where $\alpha$ is used to control the guidance weighting. Our experiments show that setting $\alpha=0.3$ usually results in a suitable \rencam{fidelity} without introducing too many hallucinated contents.

\begin{figure*}
    \centering
    \includegraphics[width=1\linewidth]{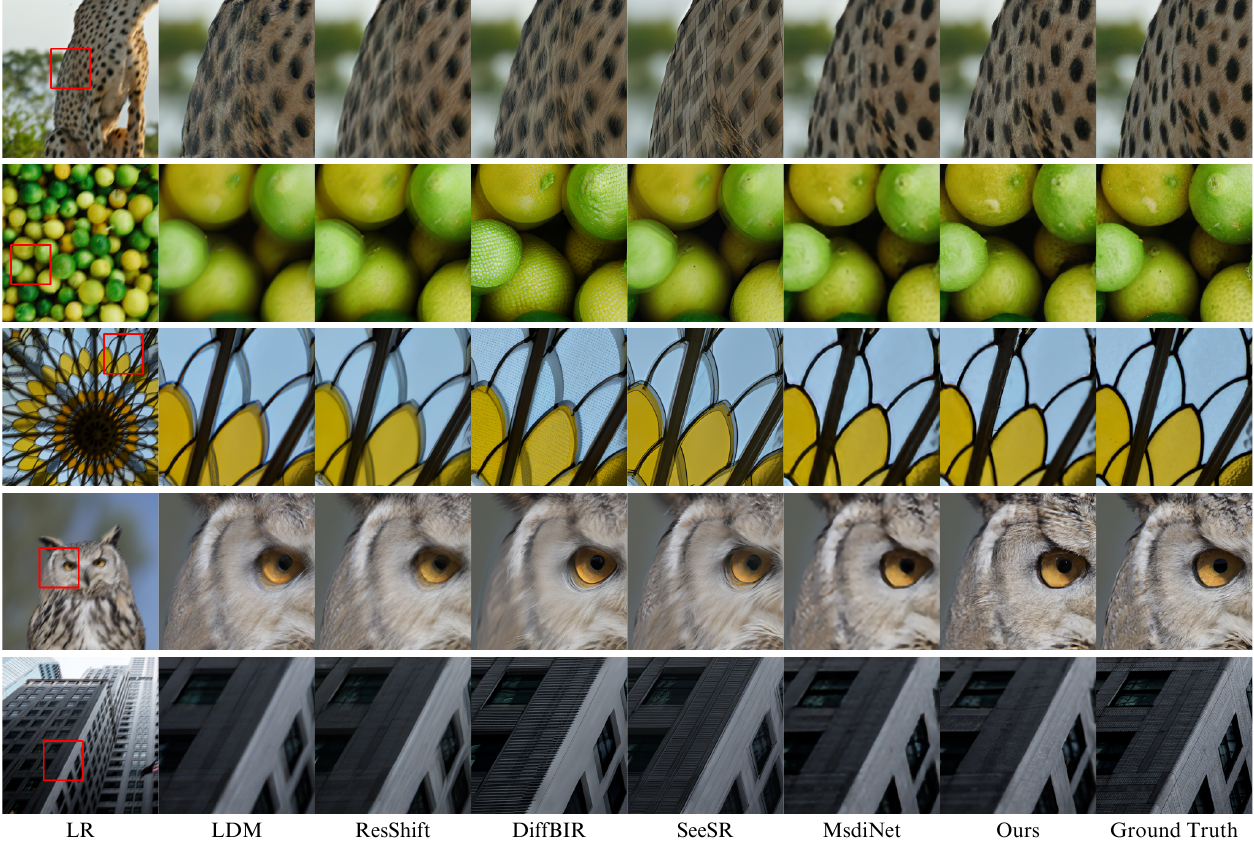}
    \caption{\heading{Qualitative comparison of $\bf{4 \times}$ upsampling on \emph{DIV2K-Val}.} The magnified areas are indicated with red boxes.}
    \label{fig:div2k_qualitative}
\end{figure*}

\section{Experiments}
\subsection{Experiment Settings}
\label{sec:settings}
\heading{Data preparation.} We evaluate our method on DIV2K unknown degradation dataset \cite{Agustsson_2017_CVPR_Workshops} and CelebA-HQ dataset~\cite{liu2015faceattributes}.


To train our degradation and restoration models, we generate LR-HR paired data as described in Supplementary. For testing dataset, We use the official train, test split for DIV2K, following SeeSR \cite{wu2024seesr} we randomly crop 1K HR patches (resolution: $256 \times 256$) and corresponding LR patches (resolution: $64 \times 64$) from the DIV2K unknown degradation validation set. We name this dataset as \emph{DIV2K-Val}. For the CelebA-HQ datasets, we randomly sample 1K images as testing data and generate corresponding LR images using the same pipeline as that for generating training data, naming this dataset \emph{CelebA-Val}. We experiment on a $4 \times$ degradation scale on \emph{DIV2K-Val}, as only $4 \times$ degraded scale LR images are provided officially. For \emph{CelebA-Val}, we experiment with both $4 \times$ and $8 \times$ degraded scales.

\begin{figure*}[ht]
    \centering
    \includegraphics[width=1\linewidth]{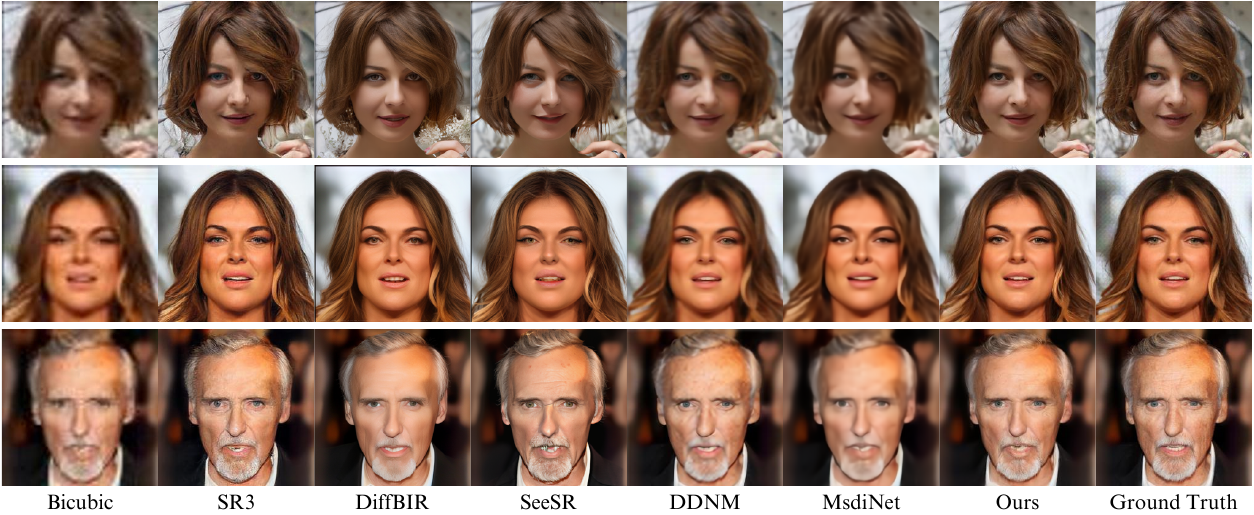}
    \caption{\heading{Qualitative comparison of $\bf{4 \times}$ upsampling on \emph{CelebA-Val}.}}
    \label{fig:celeba_qualitative}
\end{figure*}

\heading{Implementation details.} For the \emph{DIV2K-Val} testing set, we use the $256 \times 256$ uncondition denoising model pretrained on ImageNet by Dhariwal \etal \cite{dhariwal2021diffusion}. For the CelebA-Val testing set, we use $256 \time 256$ uncondition denoising model pretrained on CelebA release by Meng \etal \cite{meng2021sdedit}. For training the degradation and restoration models, we follow the settings of Li \etal \cite{li2022learning} with the addition of a consistency loss, where the learning rate of the consistency loss is set to $0.1$ of the pixel loss. During inference with the DDNM algorithm, we use DDPM as the diffusion sampling formula and adopt the spaced DDPM sampling \cite{nichol2021improved} with $100$ timesteps. The guidance scalar is set to $0.3$ throughout all the experiments.

\heading{Methods in comparison.} We compare our method with several state-of-the-art generative \rencam{model}-based blind super-resolution (SR) methods. For the \emph{DIV2K-Val} dataset, we compare our approach with GAN-based methods,  including BSRGAN \cite{zhang2021designing}, Real-ESRGAN \cite{wang2021real}, LDL \cite{liang2022details}, FeMaSR \cite{chen2022real} and DASR \cite{liang2022efficient}. We also compare with diffusion-based methods, including LDM \cite{rombach2022high}, ResShift \cite{yue2024resshift}, PASD \cite{yang2023pixel}, DiffBIR \cite{lin2023diffbir} and SeeSR \cite{wu2024seesr}. For the \emph{CelebA-Val} dataset, we compare with DiffBIR \cite{lin2023diffbir}, SeeSR \cite{wu2024seesr} and SR3 \cite{saharia2022image}, where SR3 is trained with the same training dataset we use in this work. Since our method is strongly related to \cite{wang2022zero} and MsdiNet \cite{li2022learning}, we also compare it with them. It is worth mentioning that for blind-SR tasks, the ground truth degradation kernel is unknown. Therefore, we use the default SR settings in DDNM experiments, where $\mathbf{A}$ is set to average pooling and $\mathbf{A}^\dagger$ is set to patch upsample. We use the publicly released codes and pretrained models (except SR3) of the competing methods for testing.

We employ several evaluation metrics to showcase our model's capability in terms of fidelity and \rencam{perceptual quality}. For fidelity measures, we use PSNR and SSIM \cite{wang2004image}. For perceptual quality measures, we use LPIPS \cite{zhang2018unreasonable} and DISTS \cite{ding2020image}. Additionally, we use FID \cite{heusel2017gans} to evaluate the distance between the distributions of original and restored images. These metrics provide a comprehensive evaluation of our model's performance.

\subsection{Comparison with State-of-the-Arts}
\heading{Quantitative comparisons.} We first present the quantitative comparison on \emph{DIV2K-Val} in \cref{tab:div2k}. The following observations can be made: 
1) Compared to DDNM, where $\mathbf{A}$ is set to average pooling and $\mathbf{A}^\dagger$ is set to patch upsample, our method excels in all metrics, showcasing the effectiveness of $G_d, G_r$, input perturbation, and guidance scalar.
2) Our method achieves the best scores on LPIPS, DISTS, and FID metrics, indicating that we effectively utilize the capabilities of pretrained diffusion models to produce outputs with high perceptual quality. Additionally, we achieve the best PSNR and SSIM scores compared to other generative \rencam{model}-based blind-SR methods, successfully addressing the challenge of the fidelity weakness often associated with generative \rencam{model}-based blind-SR methods. 
3) While MsdiNet has the best scores on fidelity metrics, it performs poorly on perceptual measurements. This is due to its DNN regression-based fidelity-focused learning objectives, which fall short in striking a balance between fidelity and \rencam{perceptual quality}. 

 We then present the quantitative comparison on \emph{CelebA-Val} in \cref{tab:quantitative}. Our method achieves the best scores in image perceptual measurements. We believe our method strikes a better balance between fidelity and \rencam{perceptual quality} compared to DNN regression-based methods (MsdiNet) or other generative \rencam{model}-based blind-SR methods. Overall, compared with other generative-based blind-SR methods, our method achieves better results not only in fidelity but also in image perceptual measurements.



\heading{Qualitative comparisons.} \cref{fig:div2k_qualitative} presents qualitative comparisons between our method and other diffusion-based blind-SR methods on \emph{DIV2K-Val} dataset. LDM and ResShift produce blurry results since they rely solely on the condition of low-resolution observations. DiffBIR and SeeSR can produce outputs with better image quality but often generate incorrect textures due to hallucinations from the diffusion prior. PASD results are blurry due to the challenge of correctly extracting semantic prompts from low-resolution patches. In comparison, our method, with the help of additional degradation and restoration models, can produce outputs that excel in both fidelity and \rencam{perceptual} quality.

\cref{fig:celeba_qualitative} presents qualitative comparisons between our method and other methods on the \emph{CelebA-Val} dataset. SR3 generates excessive high-frequency information. DiffBIR and SeeSR exhibit shortcomings in fidelity. DDNM, without ground truth $\mathbf{A}$ $\mathbf{A}^\dagger$, and MsdiNet struggle to generate high-quality images. In contrast, our method, with the assistance of degradation and restoration models, achieves high fidelity and \rencam{perceptual} quality.

\begin{figure*}[ht]
    \centering
    \includegraphics[width=1\linewidth]{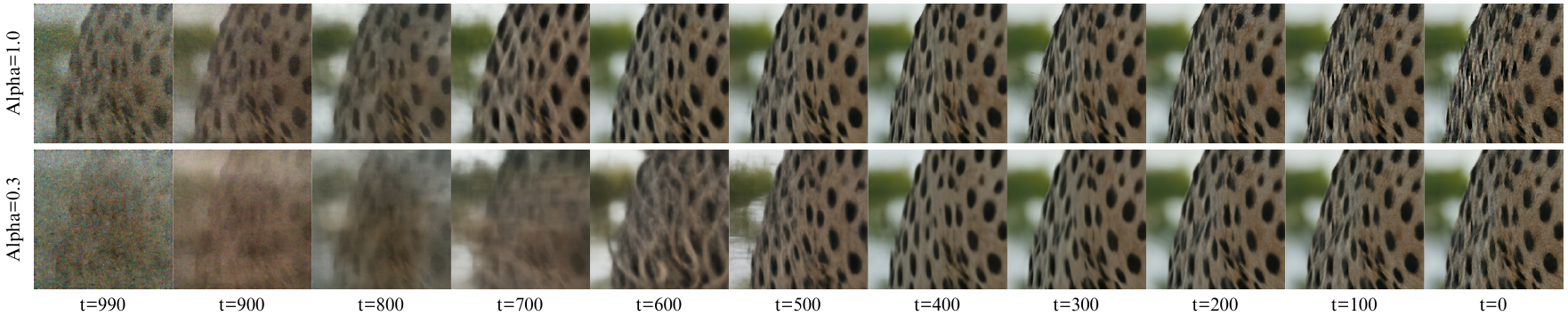}
    \caption{\heading{Ablation study on the speed of restoration process under different guidance scalar values.}}
    \label{fig:alpha_visual}
\end{figure*}

\begin{figure}
    \centering
    \includegraphics[width=1\linewidth]{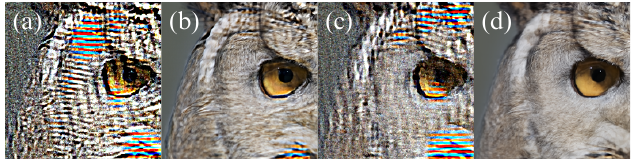}
    \caption{\heading{Ablation study on input perturbation and guidance scalar.} (a) both off, (b) guidance scalar on, (c) input perturbation on, (d) both on.}
    \label{fig:abla_visual}
\end{figure}

\subsection{Ablation Studies}
\heading{Effectiveness of input perturbation and guidance scalar.} To demonstrate the impact of input perturbation and the guidance scalar, we conducted detailed ablation studies on our proposed method using the \emph{DIV2K-Val} dataset. \shlu{The results are presented quantitatively in \cref{tab:ablation} and qualitatively in \cref{fig:abla_visual}, both of which highlight the impact of each component.}
We compared the results of simply combining DDNM \cite{wang2022zero} with the extra degradation and restoration models against incorporating input perturbation, the guidance scalar, or both. \rencam{In \cref{tab:ablation}, }\shlu{the guidance scalar significantly benefits both fidelity and perceptual quality, improving PSNR by 7.2 dB and LPIPS by 0.27. The input perturbation also benefits fidelity and perceptual quality. Combining both techniques further amplifies these gains, highlighting their effectiveness in improving our method's performance.} \rencam{\cref{fig:abla_visual} supports the same conclusion quantitatively.}

\begin{table}[]
\begin{adjustbox}{max width=\textwidth}
\begin{tabular}{cc|ccc}
\hline
\begin{tabular}[c]{@{}c@{}}Input \\ Perturbation\end{tabular} & \begin{tabular}[c]{@{}c@{}}Guidance \\ Scalar\end{tabular} & PSNR ↑                         & SSIM ↑                        & LPIPS ↓                       \\ \hline
\checkmark                                                              & \checkmark                                                           & {\color[HTML]{FF0000} 24.4681} & {\color[HTML]{FF0000} 0.6208} & {\color[HTML]{FF0000} 0.3069} \\
-                                                             & \checkmark                                                           & {\color[HTML]{0000FF} 20.7559} & {\color[HTML]{0000FF} 0.5525} & {\color[HTML]{0000FF} 0.3544} \\
\checkmark                                                              & -                                                          & 17.5641                        & 0.3585                        & 0.5147                        \\
-                                                             & -                                                          & 13.5803                        & 0.2281                        & 0.6269                        \\ \hline
\end{tabular}
\end{adjustbox}
\caption{\heading{Ablation study on input perturbation and guidance scalar.}} 
\label{tab:ablation}
\end{table}

\begin{table}[]
\centering
\begin{adjustbox}{max width=\textwidth}
\begin{tabular}{c|ccc}
\hline
Method    & PSNR ↑   & SSIM ↑  & LPIPS ↓  \\ \hline
$\alpha=0.1$ & {\color[HTML]{0000FF}24.2709} & {\color[HTML]{0000FF}0.5961} & 0.3892 \\
                              $\alpha=0.3$ & {\color[HTML]{FF0000}24.4681} & {\color[HTML]{FF0000}0.6208} & {\color[HTML]{FF0000}0.3069} \\
                              $\alpha=0.5$ & 22.387  & 0.5648 & {\color[HTML]{0000FF}0.341}  \\ 
                              $\alpha=1.0$ & 17.5641  & 0.3585 & 0.5147  \\ \hline

\end{tabular}
\end{adjustbox}
\caption{\heading{Ablation study on different values of guidance scalar.}}
\label{tab:scalar_ablation}
\end{table}

\heading{Design choices of the degradation operator and its pseudo-inverse.} We conduct experiments using various approaches to model the degradation operator and its corresponding pseudo-inverse.
We train a DNN-based degradation model and a DNN-based restoration model to \rencam{replace} the $\mathbf{A}$ and $\mathbf{A}^\dagger$ in the DDNM \cite{wang2022zero} algorithm, providing a more generalized approach to modeling the degradation and its corresponding pseudo-inverse. In our ablation experiments, we refer to this method as \emph{implicit}, as the degradation representation is implicitly learned as a feature.

Another naive approach to modeling the degradation operator and its pseudo-inverse is by defining $\mathbf{A}$ as a convolution with an explicit Gaussian kernel. This method assumes $\mathbf{A}$ is linear and known, allowing $\mathbf{A}^\dagger$ to be calculated using techniques like Singular Value Decomposition (SVD), similar to what DDNM proposes. However, relying on the assumption that $\mathbf{A}$ is a convolution operator limits its expressiveness compared to methods that learn the degradation implicitly or through more complex models. To construct an experiment based on this approach, we trained an explicit kernel estimator using the same training data described in \cref{sec:settings}. Given a low-resolution input observation, the explicit kernel estimator predicts a corresponding explicit linear degradation kernel. We define convolution with the predicted kernel as $\mathbf{A}$ and calculate its corresponding pseudo-inverse operator $\mathbf{A}^\dagger$ using SVD. In our ablation study, we refer to this method as \emph{explicit} as it explicitly defines the degradation process using a Gaussian kernel. To further experiment with the \rencam{design choices} of the $\mathbf{A}$ and $\mathbf{A}^\dagger$ in the DDNM \cite{wang2022zero} algorithm, we also explore combining the explicit version of $\mathbf{A}$ with the implicit version of $\mathbf{A}^\dagger$. In our ablation study, we refer to this approach as \emph{combine}. More details of the \emph{explicit} and \emph{combine} approaches are described in Supplementary.

We conduct the ablation study on \emph{DIV2K-Val} dataset as shown in \cref{tab:differentA}, we observe that the implicit version of $\mathbf{A}$ $\mathbf{A}^\dagger$ outperforms the \emph{combine} and \emph{explicit} approaches. This superiority is attributed to the greater expressiveness of the implicit degradation representation compared to convolution with a Gaussian kernel. Convolution with a Gaussian kernel assumes a global degradation operator, which limits its ability to capture variations in degradation that can occur locally across the image in real-world scenarios. In contrast, our \emph{implicit} approach can capture the 2D spatially varying degradation in a latent space. 

Overall, the \emph{implicit} setting of $\mathbf{A}$ and $\mathbf{A}^\dagger$ offers better expressiveness in modeling degradation, which makes our method more robust and reliable across diverse datasets and real-world scenarios.

\begin{table}[]
\centering
\begin{adjustbox}{max width=\textwidth}
\begin{tabular}{c|ccc}
\hline
Method   & PSNR ↑                           & SSIM ↑                          & LPIPS ↓                        \\ \hline
                                                                             \emph{implicit} & {\color[HTML]{FF0000} 24.4681} & {\color[HTML]{FF0000} 0.6208} & {\color[HTML]{FF0000} 0.3069} \\
                                                                             \emph{explicit} & 20.955                         & 0.3999                        & 0.5985                        \\
\emph{combine}  & {\color[HTML]{0000FF}22.4154}                        & {\color[HTML]{0000FF}0.531}                         & {\color[HTML]{0000FF}0.4787}                        \\ \hline

\end{tabular}
\end{adjustbox}
\caption{\heading{Ablation study on various approaches to modeling degradation.} All of the experiments are conducted with $4 \times$ scale.}
\label{tab:differentA}
\end{table}


\heading{Ablation study on the value of guidance scalar.} Note that the guidance scalar $\alpha$ in \cref{eq:alpha_guidance}
 is a \rencam{critical} parameter. If $\alpha$ is set too high, it accelerates the restoration process, leading to a hallucination problem in the pretrained diffusion model. Conversely, setting it too low reduces the influence of the guidance, leading to low-fidelity results. Note that setting it to zero equates to unconditional diffusion sampling.
 
 \rencam{We provide quantitative comparison in \cref{tab:scalar_ablation} and qualitative comparison in \cref{fig:alpha_visual}, both of which demonstrate the speed of restoration process under different scalar values.}
 \rencam{Both \cref{tab:scalar_ablation} and \cref{fig:alpha_visual} show that} by introducing the guidance scalar, we effectively slow down the restoration process, thereby mitigating the performance decrease caused by restoration acceleration. 

\section{Conclusions}
We have proposed a novel diffusion-based blind super-resolution (SR) solution. By combining the the degradation-aware models with DDNM, we boost diffusion guidance to produce high quality images without sacrificing the fidelity. We propose the use of a more flexible and general form of the $\mathbf{A}$ and $\mathbf{A}^{\dagger}$ in DDNM, extending their roles to \rencam{not only degradation but also restoration}. The integration of input perturbation and guidance scalar further enhances the performance of our method, reducing the impact of degradation estimation errors and improving the synergy between our models and DDNM. Through these innovations, we demonstrate that our method excels in both fidelity and perceptual quality compared to state-of-the-art diffusion-based blind SR methods.


{\small
\bibliographystyle{ieee_fullname}
\bibliography{egbib}
}

\end{document}